\documentclass[preprint,aps,floats]{revtex4}
\usepackage{epsfig}

\usepackage{graphicx}
\usepackage{dcolumn}
\usepackage{bm}
\def\lsim{\mathrel {\vcenter {\baselineskip 0pt \kern 0pt
    \hbox{$&lt;$} \kern 0pt \hbox{$\sim$} }}}
\def\gsim{\mathrel {\vcenter {\baselineskip 0pt \kern 0pt
    \hbox{$&gt;$} \kern 0pt \hbox{$\sim$} }}}

\newcommand{\U}{{\cal {U}}}

\begin{document}

\title{Interactions of Unparticles with Standard Model Particles}

\author{Shao-Long Chen}
\email{shaolong@phys.ntu.edu.tw}
\author{Xiao-Gang He}
\email{hexg@phys.ntu.edu.tw}
\affiliation{ Department of Physics
and Center for Theoretical Sciences, National Taiwan University,
Taipei, Taiwan}

\date{\today}

\begin{abstract}
We study interactions of unparticles ${\cal {U}}$ of dimension
$d_\U$ due to Georgi with Standard Model (SM) fields through
effective operators. The unparticles describe the low energy
physics of a non-trivial scale invariant sector. Since unparticles
come from beyond the SM physics, it is plausible that they
transform as a singlet under the SM gauge group. This helps
tremendously in limiting possible interactions. We analyze
interactions of scalar ${\cal {U}}$, vector ${\cal {U}}$$^\mu$ and
spinor ${\cal {U}}$$^s$ unparticles with SM fields and derivatives
up to dimension four. Using these operators, we discuss different
features of producing unparticles at $e^+ e^-$ collider and other
phenomenologies. It is possible to distinguish different
unparticles produced at $e^+e^-$ collider by looking at various
distributions of production cross sections.
\end{abstract}


\maketitle

In a scale invariant theory in four space-time dimensions, there
are no particles with a non-zero mass. In our real world, there
are plenty of particles with non-zero masses. If scale invariance
plays a role in nature, it must have been broken at some high
energy scale beyond the Standard Model (SM) scale. At low energy,
our world is described by the SM. At high energy there may be both
scale invariant sector and also other sectors which do not have
scale invariant such as the SM fields. Recently Georgi proposed an
interesting idea to describe possible scale invariant effect at
low energies, termed unparticle\cite{Georgi:2007ek}. Based on a
specific model scale invariant theory by Banks and Zaks
\cite{Banks:1982gt}, Georgi argued that operators $O_{BZ}$ made of
BZ fields may interact with operators $O_{SM}$ made of SM fields
at some high energy scale by exchange particles with large masses,
$M_{\cal{U}}$, with the generic form $O_{SM}
O_{BZ}/M^k_{\cal{U}}$. At another scale $\Lambda_{\cal{U}}$ the BZ
sector induce dimensional transmutation, below that scale the BZ
operator $O_{BZ}$ matches on to unparticle operator $O_{\cal{U}}$
with dimension $d_\U$ and the unparticle interaction with SM
particles at low energy has the form $\lambda \Lambda_{\cal{U}}
^{4-d_{SM} - d_\U} O_{SM} O_{\cal{U}}$. Here $d_{SM}$ is the
dimension of the operator $O_{SM}$.

An unparticle looks like a non-integral $d_\U$ dimension invisible
particle. Depending on the nature of the original operator
$O_{BZ}$ and the transmutation, the resulting unparticles may have
different Lorentz structure. We will indicate an unparticle acts
like a Lorentz scalar as $O_{\cal{U}}$, a vector as
$O_{\cal{U}}^\mu$ and a spinor as $O^s_{\cal{U}}$. If all
interactions are perturbative, one maybe able to calculate the
dimension $d_\U$ and also the coupling $\lambda$. But the matching
from the BZ physics to the unparticle physics will be a
complicated strong interaction problem to deal with. One can work
with the effective coupling $\lambda$ for phenomenology which has
been practiced by many.

For detailed studies, one also needs to know how an unparticle
interacts with SM particles. Recent studies have focused on
several low dimension
operators\cite{Georgi:2007ek,Georgi:2007si,Cheung:2007ue,
Luo:2007bq,Liao:2007bx,Ding:2007bm,Chen:2007vv,
Aliev:2007qw,Li:2007by,
Lu:2007mx,Fox:2007sy,Stephanov:2007ry,
Greiner:2007,Davoudiasl:2007,Choudhury:2007}. The unparticle
interactions with the SM particles are through exchange of some
other heavy particles of mass $M_{\cal{U}}$. Therefore the form of
the interaction is basically determined by the nature of the heavy
particle. If it is a SM singlet, the unparticle $O_{\cal{U}}$
resulting from the transmutation should also transform under the
SM gauge group as a singlet. One cannot rule out other
possibilities. There are many ways that the SM fields can couple
to an unparticle. If the unparticle is a SM singlet, the
possibilities are limited since the SM fields have to form SM
singlet first. In this work we concentrate on possible
interactions of unparticles with the SM particles assuming that
unparticles transform as SM singlets and study some implications.

In the following we list operators composed of SM fields and
derivatives with dimensions less than or equal to 4 invariant
under the SM gauge group.
\begin{eqnarray}
\mbox{Scalar}&&\mbox{ $O_{\cal{U}}$ couplings}: \nonumber\\
a)&& \mbox{Couplings with gauge bosons}\nonumber\\
 &&\lambda_{gg}
\Lambda_{\cal{U}}^{-d_\U}G^{\mu\nu}G_{\mu\nu}O_{\cal{U}},\;\lambda_{ww}
\Lambda_{\cal{U}}^{-d_\U}W^{\mu\nu}W_{\mu\nu}O_{\cal{U}},
\;\lambda_{bb}\Lambda^{-d_\U}_{\cal{U}}B^{\mu\nu}B_{\mu\nu}O_{\cal{U}},\nonumber\\
&&\tilde \lambda_{gg} \Lambda_{\cal{U}}^{-d_\U}\tilde
G^{\mu\nu}G_{\mu\nu}O_{\cal{U}}, \;\tilde \lambda_{ww}
\Lambda_{\cal{U}}^{-d_\U}\tilde W^{\mu\nu}W_{\mu\nu}O_{\cal{U}},
\;\tilde \lambda_{bb}\Lambda^{-d_\U}_{\cal{U}}\tilde B^{\mu\nu}B_{\mu\nu}O_{\cal{U}},\\
b) &&\mbox{Coupling with Higgs and Gauge bosons}\nonumber\\
 &&\lambda_{hh}
\Lambda_{\cal{U}}^{2-d_\U}H^\dagger H O_{\cal{U}},\;\tilde
\lambda_{hh}\Lambda_{\cal{U}}^{-d_\U}(H^\dagger D_\mu H)
\partial^\mu O_{\cal{U}}\;,\nonumber\\
&&\;\lambda_{4h}\Lambda_{\cal{U}}^{-d_\U}(H^\dagger
H)^2O_{\cal{U}},
\;\lambda_{dh}\Lambda_{\cal{U}}^{-d_\U} (D_\mu H)^\dagger (D^\mu H)O_{\cal{U}},\\
c)&&\mbox{Couplings with fermions and gauge bosons}\nonumber\\
&&\lambda_{QQ}\Lambda_{\cal{U}}^{-d_\U} \bar Q_L \gamma_\mu D^\mu
Q_L O_{\cal{U}}, \;\lambda_{UU}\Lambda_U^{-d_\U}\bar U_R\gamma_\mu
D^\mu U_R O_{\cal{U}},
\;\lambda_{DD}\Lambda_{\cal{U}}^{-d_\U}\bar D_R \gamma_\mu D^\mu D_R O_{\cal{U}},\nonumber\\
&&\lambda_{LL}\Lambda_{\cal{U}}^{-d_\U}\bar L_L \gamma_\mu D^\mu
L_L O_{\cal{U}}, \;\lambda_{EE}\Lambda_{\cal{U}}^{-d_\U}\bar
E_R\gamma_\mu D^\mu E_R O_{\cal{U}},
\;\lambda_{\nu\nu}\Lambda_{\cal{U}}^{-d_\U}\bar \nu_R\gamma_\mu D^\mu \nu_R O_{\cal{U}},\nonumber\\
&&\tilde \lambda_{QQ}\Lambda_{\cal{U}}^{-d_\U}\bar Q_L \gamma_\mu
Q_L
\partial^\mu O_{\cal{U}}, \;\tilde \lambda_{UU}\Lambda_{\cal{U}}^{-d_\U}\bar U_R \gamma_\mu
U_R \partial^\mu O_{\cal{U}},
\;\tilde \lambda_{DD}\Lambda_{\cal{U}}^{-d_\U}\bar D_R  \gamma_\mu D_R \partial^\mu O_{\cal{U}},\nonumber\\
&&\tilde \lambda_{LL}\Lambda_{\cal{U}}^{-d_\U}\bar L_L  \gamma_\mu
L_L
\partial^\mu O_{\cal{U}}, \;\tilde \lambda_{EE}\Lambda_\U^{-d_\U}\bar E_R \gamma_\mu E_R
\partial^\mu O_{\cal{U}},
\;\tilde \lambda_{RR}\Lambda_{\cal{U}}^{-d_\U}\bar \nu_R \gamma_\mu \nu_R \partial^\mu O_{\cal{U}},\nonumber\\
&&\lambda_{YR}\Lambda_{\cal{U}}^{1-d_\U}\bar \nu_R^C \nu_R O_{\cal{U}},\\
d)&&\mbox{Couplings with fermions and Higgs boson}\nonumber\\
&&\lambda_{YU}\Lambda_{\cal{U}}^{-d_\U}\bar Q_L H U_R O_{\cal{U}},
\;\lambda_{YD}\Lambda_{\cal{U}}^{-d_\U}\bar Q_L \tilde H D_R O_{\cal{U}},\nonumber\\
&&\lambda_{Y\nu}\Lambda_{\cal{U}}^{-d_\U}\bar L_L
H\nu_RO_{\cal{U}},\;\lambda_{YE}\Lambda_{\cal{U}}^{-d_\U}\bar L_L
\tilde H E_R O_{\cal{U}}, \\
\mbox{Vector}&&\mbox{ $O^\mu_{\cal{U}}$ couplings}:\nonumber\\
a)&&\mbox{Couplings with fermions}\nonumber\\
&&\lambda'_{QQ}\Lambda_{\cal{U}}^{1-d_\U}\bar Q_L \gamma_\mu Q_L
O^\mu_{\cal{U}}, \;\lambda'_{UU}\Lambda_{\cal{U}}^{1-d_\U}\bar U_R
\gamma_\mu U_R O_{\cal{U}}^\mu,
\;\lambda'_{DD}\Lambda_{\cal{U}}^{1-d_\U}\bar D_R  \gamma_\mu D_R O^\mu_{\cal{U}},\nonumber\\
&&\lambda'_{LL}\Lambda_{\cal{U}}^{1-d_\U}\bar L_L  \gamma_\mu L_L
O^\mu_{\cal{U}}, \;\lambda'_{EE}\Lambda_{\cal{U}}^{1-d_\U}\bar E_R
\gamma_\mu E_R O^\mu_{\cal{U}},
\;\lambda'_{RR}\Lambda_{\cal{U}}^{1-d_\U}\bar \nu_R \gamma_\mu \nu_R O_{\cal{U}}^\mu,\nonumber\\
b)&& \mbox{Couplings with Higg boson and Gauge bosons}\nonumber\\
&&\lambda'_{hh}\Lambda_{\cal{U}}^{1-d_\U}(H^\dagger D_\mu H)
O^\mu_{\cal{U}}\;,\;\lambda'_{bO}\Lambda_{\cal{U}}^{1-d_\U}B_{\mu\nu}\partial^\mu O^\nu\;.\\
\mbox{Spinor}&&\mbox{ $O^s_{\cal{U}}$ couplings}:\nonumber\\
&&\lambda_{s\nu} \Lambda_{\cal{U}}^{5/2-d_\U} \bar \nu_R
O^s_{\cal{U}},\;\lambda_s \Lambda_{\cal{U}}^{3/2-d_\U}\bar L_L H
O^s_{\cal{U}}\;.
\end{eqnarray}
Here $G$, $W$ and $B$ are the $SU(3)_C$, $SU(2)_L$ and $U(1)_Y$
gauge fields, respectively. $Q_L$, $U_R$, $D_R$, $L_L$, $E_R$ are
the SM left-handed quark doublet, right-handed up-quark,
right-handed down-quark, left-handed lepton doublet and
right-handed charged lepton, respectively. In the above we also
included the right handed neutrino $\nu_R$ which might be needed from
neutrino oscillation data.

The scalar $\cal{U}$ unparticle has the largest number of
operators. In this class of interactions, the lowest SM dimension
operators is the coupling of $\cal{U}$ to two Higgs fields,
$H^\dagger H O_{\cal{U}}$. The second lowest operator involves two
right handed neutrinos, $\bar \nu_R^C \nu_R O_{\cal{U}}$.  The
rest have the same dimensions with the SM fields and derivatives
forming dimension four operators. In the following we point out
some interesting features.

The operator $H^\dagger H O_{\cal{U}}$ with a low dimension
$\Lambda_{\cal{U}}^{2-d_\U}$ may have the best chance to show up
at low energies. An effect is that when the Higgs field develops a
non-zero vacuum expectation value (vev) $\langle H\rangle =
v/\sqrt{2}$ as required by gauge symmetry breaking and generation
of SM particle masses, there is a tadpole coupling $\lambda_{hh}
\Lambda_{{\cal U}}^{2-d_\U} v^2/2$ of unparticle to vev which
introduces a scale to the unparticle sector. This interaction will
cause the unparticle sector to be pushed away from its scale
invariant fixed point and the theory become non-scale invariant at
some low scale. Below that the unparticle sector presumably
becomes a traditional particle sector\cite{Fox:2007sy}. We note
that this may not be necessarily true if one also include the
other operator $(H^\dagger H)^2O_{\cal{U}}$. This term also has a
tadpole coupling of unparticle to vev. It is given by
$\lambda_{4h} \Lambda_{\cal U }^{-d_\U}$. If $\lambda_{hh}
\Lambda_{{\cal U}}^2 + \lambda_{4h} v^2/2 = 0$, the tadpole will
be removed. One then has
\begin{eqnarray}
\lambda_{hh} \Lambda_{{\cal U}}^{2-d_\U}H^\dagger H O_{\cal{U}}  +
\lambda_{4h} \Lambda_{\cal U }^{-d_\U}(H^\dagger H)^2 O_{\cal{U}} =
{1\over 4}\lambda_{4h} \Lambda_{\cal U }^{-d_\U}(h^4 + 4 v h^3 + 5
v^2h^2 + 2 v^3 h)O_{\cal U}.
\end{eqnarray}
Here, we have removed the would-be Goldstone boson in the Higgs
field and $h$ is the physical Higgs field. The operators above
will induce mixing between $h$ and the scalar unparticle. A
physical Higgs may oscillate into $O_{\cal U}$ and disappear. We
should note that the cancellation mechanism discussed above is by
assumption. We are not able to find a symmetry to guarantee it and
it may not be stable. Another possibility is that these couplings
cannot be generated such that unparticle physics effect can still
show up at low energies. More studies are needed.

The operator $\bar \nu_R^C \nu_R O_{\cal{U}}$ involves
right-handed neutrino interaction $\nu_R$ with an unparticle. If
$\nu_R$ is heavy, there is no observable effect. If $\nu_R$
turns out to be a light sterile neutrino, one may see some effects
in neutrino decays, a heavier $\nu_R$ may decay into a lighter
$\nu_R$ and $O_{\cal U}$. Such effects may be difficult to
observe.

There are six operators involve $O_{\cal U}$ and gauge particles.
The interactions with gluon fields can produce $O_{\cal U}$ at
hadron colliders through $g g \to g {\cal U}$, $q\bar q \to g
{\cal U}$ and $g q \to q {\cal U}$. The operators with $W$ and $B$
can produce ${\cal U}$ at a photon collider through $\gamma \gamma
\to {\cal U}$, $\gamma e \to \gamma {\cal U}$, and also
interesting signature in $WW$ scattering\cite{Greiner:2007}.

The operators in class c) have rich phenomenology. Several of the
the operators have been studied in flavor changing decay of a
heavy fermion to a light fermion plus an unparticle such as $t \to
u(c) + {\cal U}$, meson and anti-meson mixing, and other flavor
changing decays. These operators can also produce ${\cal U}$ at
hadron and $e^+e^-$ colliders. We will come back to this later in
discussing how to distinguish different types of unparticles.

The operators in class d) involve an unpartile with SM Yukawa
terms\cite{Fox:2007sy}. These will open new decay channels for the
Higgs and the top quark with an unparticle in the final state.
They can also induce $e^+e^- (q \bar q) \to h {\cal U}$.

There are less operators involve the vector $\cal{U} ^\mu$ and SM
particle. The class a) operators are the most studied ones.
Similar to class c) operators for scalar unparticle couplings,
they can induce $t \to u(c) + {\cal U}$, meson and anti-meson
mixing and can also produce ${\cal U}$ at hadron and $e^+e^-$
colliders. The first operator in class b) can induce $h$ and
unparticle mixing, and the second operator can induce $B$ and
unparticle mixing.

The unparticle may have spinor structure under the Lorentz
group\cite{Luo:2007bq}. There are only two operators for spinor
unparticle and SM particle interactions. The operator $\bar \nu_R
O^s_{\cal{U}}$ has the lowest dimension in the whole list. This
operator will mixing the unparticle with right-handed neutrino.
Deviations of neutrino oscillation pattern may be the best place
for looking for unparticle effects. The operator $\bar L_L H
O^s_{\cal{U}}$ can induce mixing between left-handed neutrino and
an unparticle. Again this will affect neutrino mixing and also
cause the PMNS mixing matrix to be not the usual $3\times 3$
unitary matrix for three left handed neutrinos. This operator will
cause Higgs to decay into a neutrino and an unparticle.

There are a lot of interesting phenomenology which can be carried
out using the above listed interactions. Besides the production of
unpartiles, there are also virtual effects of
unparticles\cite{Georgi:2007ek,Georgi:2007si,Cheung:2007ue,Luo:2007bq,
Liao:2007bx,Ding:2007bm,Chen:2007vv, Aliev:2007qw,Li:2007by,
Lu:2007mx,Fox:2007sy,Stephanov:2007ry,
Greiner:2007,Davoudiasl:2007,Choudhury:2007}. In the rest of the
paper we concentrate on the possibility of distinguishing whether
an $O_{\cal{U}}$ or an $O^\mu_{\cal{U}}$ is produced through
$e^+e^-$ collider through $e^+e^- \to \gamma (Z) + {\cal
{U}},\;\;\gamma ( Z)+  {\cal {U}}^\mu$.

\begin{figure}[t!]
\includegraphics[width=6.0 in]{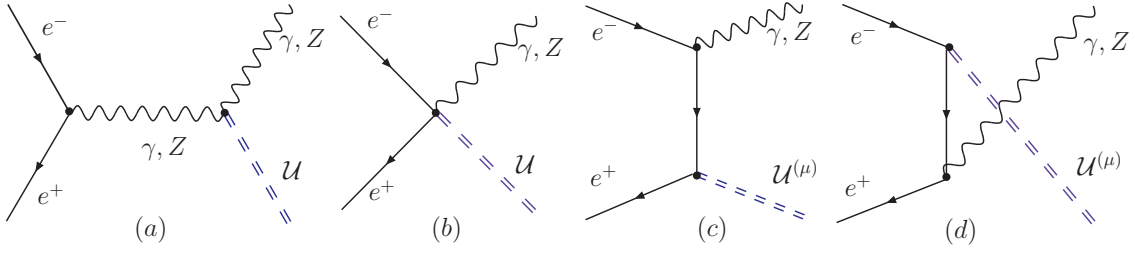}
\caption{\label{feynman} \small The Feynman diagrams for
${\cal{U}}$ and ${\cal U}^\mu$ productions through $e^+e^- \to
\gamma (Z) + {\cal U}^{(\mu)}$. For ${\cal U}^\mu$ production,
only diagrams (c) and (d) contribute.}
\end{figure}

The Feynman diagrams for the above processes are shown in Fig. 1.
For ${\cal U}$ production, the operators
$\lambda_{bb}B^{\mu\nu}B_{\mu\nu}O_{\cal{U}},\;
\lambda_{ww}W^{\mu\nu}W_{\mu\nu}O_{\cal{U}},\; \tilde
\lambda_{ww}\tilde W^{\mu\nu}W_{\mu\nu}O_{\cal{U}},\; \tilde
\lambda_{bb}\tilde B^{\mu\nu}B_{\mu\nu}O_{\cal{U}}$ contribute
through s-channel, and the operators $\lambda_{LL}\bar L_L
\gamma_\mu D^\mu L_L O_{\cal{U}}$ and $\lambda_{EE}\bar
E_R\gamma_\mu D^\mu E_R O_{\cal{U}}$ contribute through by
$u,t$-channels. The operators $\bar L_L \gamma_\mu L_L
\partial^\mu O_{\cal{U}}$ and $\bar E_R \gamma_\mu E_R
\partial^\mu O_{\cal{U}}$ do not contribute.
For ${\cal U}^\mu$ production, the operators $\lambda'_{LL}\bar
L_L \gamma_\mu L_L O^\mu_{\cal{U}}$ and $\lambda'_{EE}\bar E_R
\gamma_\mu E_R O^\mu_{\cal{U}}$ contribute through $u,t$-channels.

Carrying out the phase integral for the unparticle, we find that
for $e^+(p_1) e^-(p_2) \to \gamma(p_3) + {\cal{U}}^{(\mu)}(P_\U)$,
the cross section is given by
\begin{equation}
\frac{d \sigma}{d E_\gamma} = \frac{1}{2 s} \,  | \overline{{\cal M}} |^2 \;
 \frac{ E_\gamma A_{d_{\cal{U}}}{(P^2_\U)}^{d_\U - 2} }{ 16 \pi^3}
\, d \Omega  \;,
\end{equation}
where $|\overline{{\cal M}}|^2$ is the initial spin averaged
matrix element squared. $\Omega$ is the photon solid angle.

In the above, we have followed Ref.\cite{Georgi:2007ek} using
 $A_{d_\U} \theta(p^0_{\cal
U}) \theta(p^2_{\cal U}) (p^2_{\cal U})^{d_\U -2}$ for bosonic unparticle phase space factor,
with $A_{du} =(16 \pi^{5/2}/(2\pi)^{2d_\U})\Gamma(d_\U+1/2)/(\Gamma(d_\U - 1)\Gamma(2d_\U)$).

While for the processes $e^+  e^- \to Z + {\cal{U}}^{(\mu)}$, the
cross section is given by
\begin{equation}
\frac{d \sigma}{d E_Z} = \frac{1}{2 s} \, | \overline{{\cal M}} |^2 \;
 \frac{ \sqrt{E_Z^2-m_Z^2} A_{d_{\cal{U}}}{\left(P^2_\U\right)}^{d_\U - 2} }{ 16 \pi^3}
\,  d \Omega  \;.
\end{equation}
Here $\Omega$ is the $Z$ solid angle.

\begin{figure}[hbt!]
\begin{center}
\hskip -0.5cm \epsfig{file=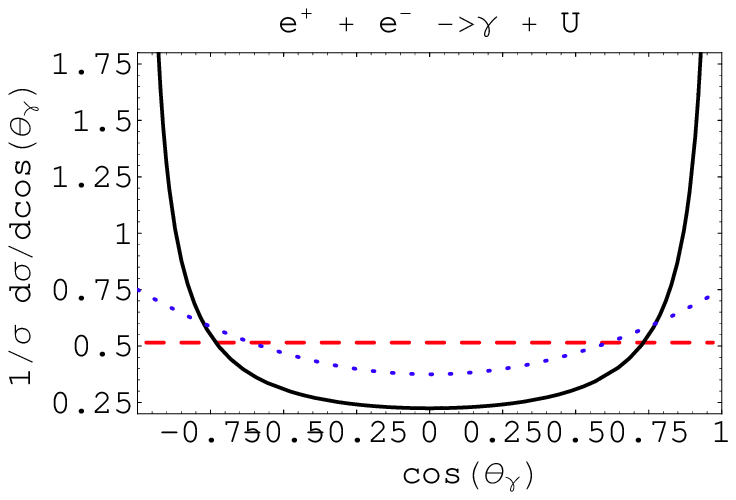,width=.45\textwidth} \hskip
-0.2cm \epsfig{file=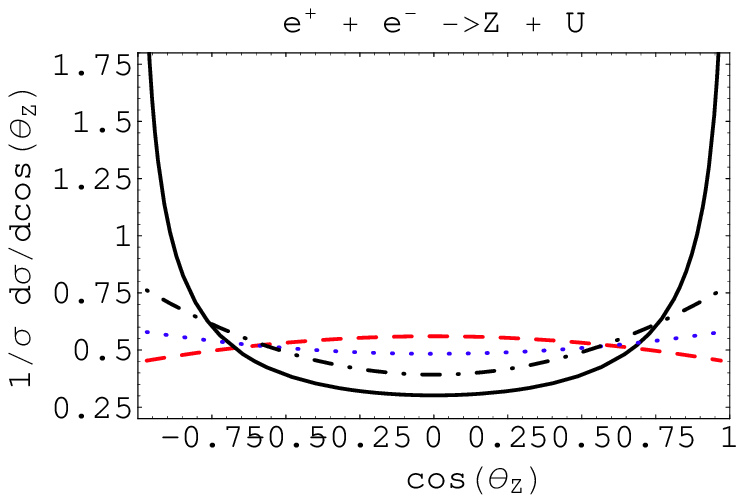,width=.45\textwidth} \vskip -0.1cm
\hskip -0.5cm \epsfig{file=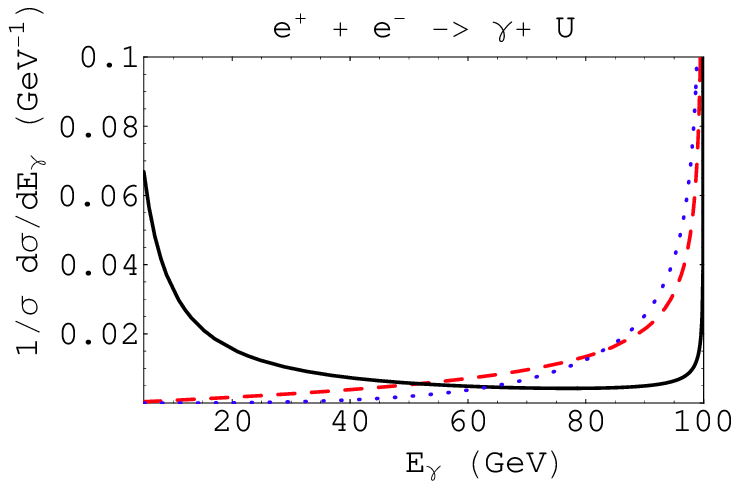,width=.45\textwidth} \hskip
-0.2cm \epsfig{file=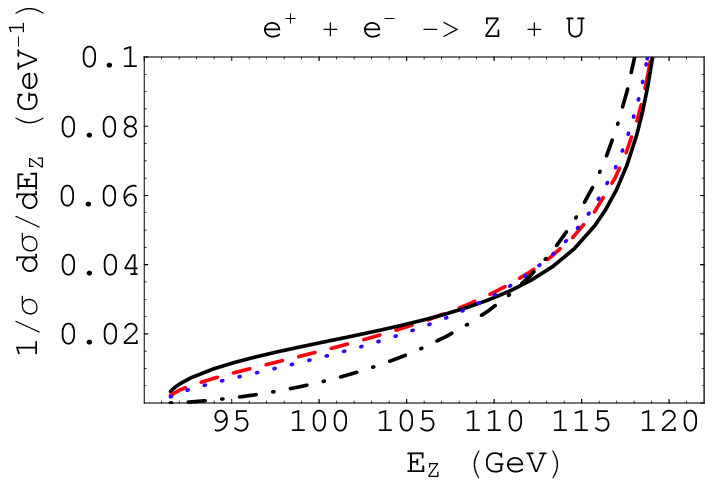,width=.45\textwidth} \caption{\small
Normalized photon (Z boson) energy spectrum and angular
distribution of $e^+e^-\to \gamma (Z)+\U$ for $d_\U=1.5$ at
$\sqrt{s}=200$ GeV. Dashed, solid, dotted and dot-dashed curves
represent the contributions from the operators with couplings
$\lambda_{LL,EE}$ ($\lambda_{LL}$ and $\lambda_{EE}$ give  same
distributions), $\lambda'_{LL,EE}$ ($\lambda'_{LL}$ and
$\lambda'_{EE}$ give same distributions), $\lambda_{ww,bb}$ and
$\tilde\lambda_{ww,bb}$, respectively. Note that for the left
panel, the curves are identical for the contributions from
$\lambda_{ww,bb},\tilde\lambda_{ww,bb}$. We have imposed
$|\cos\theta_{\gamma,Z}|<0.97$ and $E_\gamma> 5$ GeV.}
\label{fig:200GeV}
\end{center}
\end{figure}

The matrix elements squared for different operators are given by:
\begin{eqnarray}
|\overline{\cal M}|^2(e^+e^-\to\gamma \U)
&=&\frac{e^2s(u^2+t^2)}{\Lambda_\U^{2d_\U}}f(\lambda_{ww},\lambda_{bb})\;,
\nonumber \\
 |\overline{\cal M}|^2(e^+e^-\to Z \U)&=&
\frac{e^2s[(t-m_Z^2)^2+(u-m_Z^2)^2+2sm_Z^2]}
{\Lambda_\U^{2d_\U}}g(\lambda_{ww},\lambda_{bb})\;, \nonumber\\
|\overline{\cal M}|^2(e^+e^-\to\gamma \U)
&=&\frac{e^2s(u^2+t^2)}{\Lambda_\U^{2d_\U}}f(\tilde\lambda_{ww},\tilde\lambda_{bb})\;,
\nonumber \\
 |\overline{\cal M}|^2(e^+e^-\to Z \U)&=&
\frac{e^2s[(t-m_Z^2)^2+(u-m_Z^2)^2-2sm_Z^2]}
{\Lambda_\U^{2d_\U}}g(\tilde\lambda_{ww},\tilde\lambda_{bb})\;,\nonumber\\
|\overline{\cal {M}}|^2(e^+e^- \to \gamma \U) &= & \frac{ 2 e^2
(\lambda_{LL}^2+\lambda_{EE}^2)} {(\Lambda_\U^2)^{d_\U-1}}
\frac{s}{\Lambda_\U^2} \;,
\\
|\overline{\cal {M}}|^2(e^+e^- \to Z \U ) &= &
\frac{2g^2}{\cos^2\theta_w}
\frac{\left(\lambda_{LL}^2(\frac{1}{2}-\sin^2\theta_w)^2+\lambda_{EE}^2\sin^4\theta_w\right)}
{(\Lambda_\U^2)^{d_\U-1}}
\frac{(s+\frac{ut}{2m_Z^2})}{\Lambda_\U^2} \;,
\nonumber\\
|\overline{\cal {M}}|^2(e^+e^- \to \gamma {\cal {U}}^\mu) & = &
\frac{e^2(\lambda_{LL}'^2+\lambda_{EE}'^2)}{(\Lambda_\U^2)^{d_\U-1}}
\frac{u^2+t^2+2sP_\U^2}{ut} \;,
\nonumber\\
|\overline{\cal {M}}|^2(e^+e^- \to Z {\cal {U}}^\mu) &=&
\frac{g^2}{\cos^2\theta_w}
\frac{\left(\lambda_{LL}'^2(\frac{1}{2}-\sin^2\theta_w)^2+\lambda_{EE}'^2\sin^4\theta_w\right)}
{(\Lambda_\U^2)^{d_\U-1}}
\nonumber\\
&&\times\frac{\left(u^2+t^2+2s(P_\U^2+m_Z^2)-P_\U^2m_Z^2(\frac{u}{t}+\frac{t}{u})\right)}{ut}
\;, \nonumber
\end{eqnarray}
where
\begin{eqnarray}
f(\lambda_{ww},\lambda_{bb})&=&
\frac{(\lambda_{ww}\sin^2\theta_w+\lambda_{bb}\cos^2\theta_w)^2}{s^2}
\nonumber\\ &&
+\frac{(\lambda_{ww}-\lambda_{bb})^22[(\frac{1}{2}-\sin^2\theta_w)^2+\sin^4\theta_w]}
{(s-m_Z^2)^2} \nonumber\\ &&
+\frac{(\lambda_{ww}\sin^2\theta_w+\lambda_{bb}\cos^2\theta_w)(\lambda_{ww}-\lambda_{bb})
(1-4\sin^2\theta_w)}{s(s-m_Z^2)}\;,
\nonumber\\
g(\lambda_{ww},\lambda_{bb})&=&
\frac{(\lambda_{ww}-\lambda_{bb})^2\sin^2 2\theta_w}{s^2}
\\
&& +\frac{(\lambda_{ww}\cot\theta_w+\lambda_{bb}\tan\theta_w)^2
[(\frac{1}{2}-\sin^2\theta_w)^2+\sin^4\theta_w]}{2(s-m_Z^2)^2}
\nonumber\\ &&
+\frac{(\lambda_{ww}-\lambda_{bb})(\lambda_{ww}\cos^2\theta_w+\lambda_{bb}\sin^2\theta_w)
(1-4\sin^2\theta_w)}{s(s-m_Z^2)} \;.\nonumber
\end{eqnarray}

The above matrix elements squared would give very different energy
and angular distributions for $\U$ and $\U^\mu$ productions since
the $s$, $u$ and $t$ parameters appear in different combinations.
The photon and Z boson angular and energy distributions are
plotted taking $d_\U=1.5$ and $\sqrt{s}=200$ GeV, in Fig. 2,
relevant for LEP II data, and also 500 GeV, in Fig. 3, relevant for
ILC for illustration. In both Figs. 2 and 3, we plot the
distributions with different couplings setting others to zero.
Since the electron mass is small, the cross section diverges at
$\cos\theta = \pm 1$ for $\U^\mu$ production due to $u$ and $t$
appear in the denominators if electron mass is neglected, but
finite for $\U$ production. This provides a clear way to
distinguish $\U$ and $\U^\mu$ production as can be seen from
Figs. 2 and 3. The energy distributions can also provide useful
information which can also be seen from Figs. 2 and 3. For this
case we made a cut with $|\cos\theta_{\gamma, Z}| < 0.97$ to avoid
the divergence for $\U^\mu$ production at $|\cos\theta_{\gamma,
Z}| = 1$.
\begin{figure}[hbt!]
\begin{center}
\hskip -0.5cm \epsfig{file=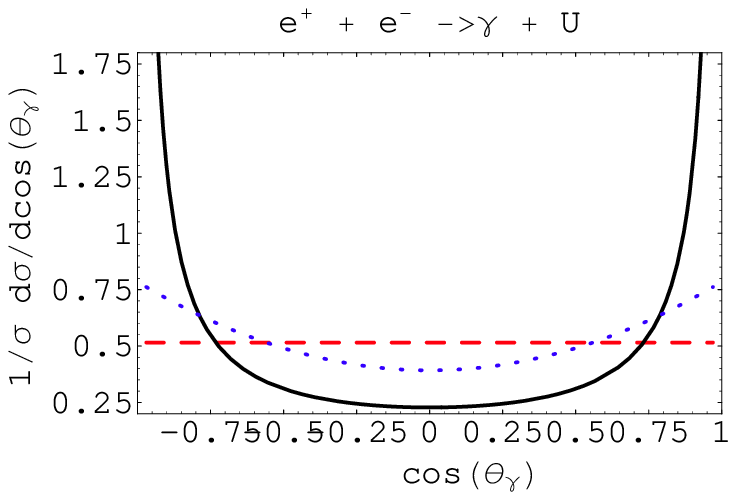,width=.45\textwidth} \hskip
-0.2cm \epsfig{file=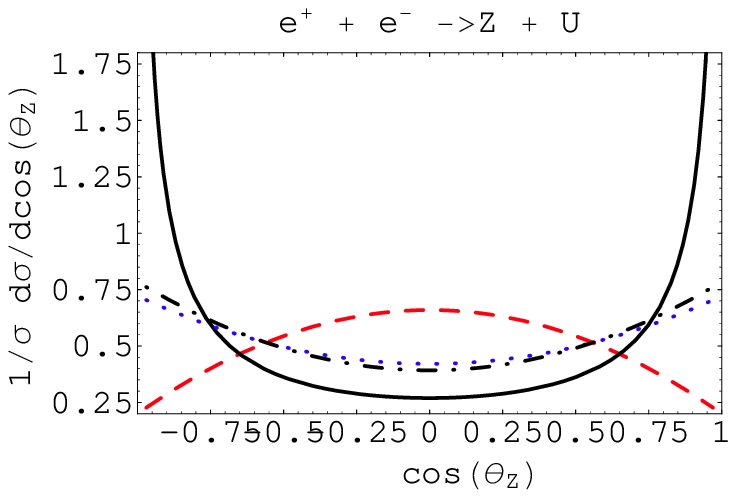,width=.45\textwidth} \vskip -0.1cm
\hskip -0.5cm \epsfig{file=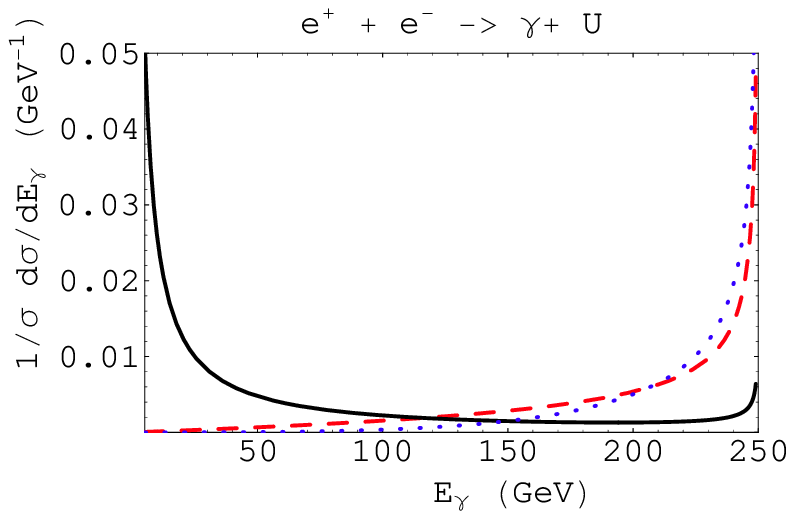,width=.45\textwidth} \hskip
-0.2cm \epsfig{file=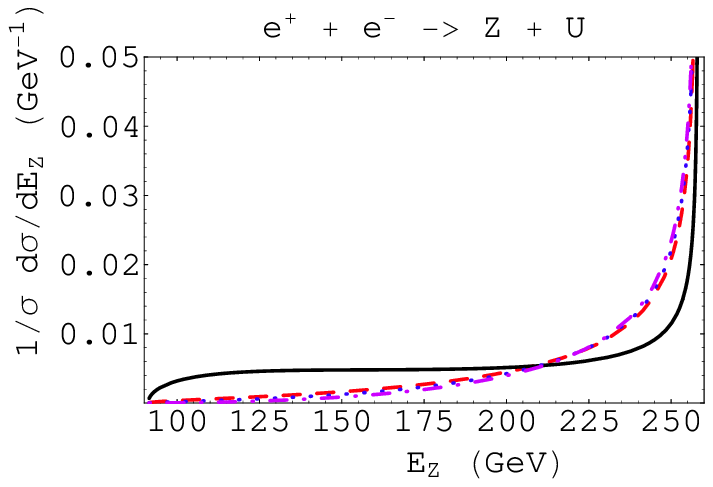,width=.45\textwidth} \caption{\small
Normalized photon (Z boson) energy spectrum and angular
distribution of $e^+e^-\to \gamma (Z)+\U$ for $d_\U=1.5$  at
$\sqrt{s}=500$ GeV.  }
\end{center}\label{fig:500GeV}
\end{figure}

The above discussion show that the study of various distributions
may be able to provide information about the type of unparticles
at ILC if a large number of events can be obtained. One needs,
however, to see if existing constraints have already ruled out
such possibilities. We now comment on constraints on the relevant
coupling of the operators.

Presently there is a direct constraint from LEP II
data~\cite{lep-ph} at $\sqrt{s} = 207$ GeV on the cross section
for $e^+ e^- \to \gamma X^0$ where $X^0$ is invisible.  With
 the cuts $E_\gamma > 5 GeV$ and $|\cos\theta| < 0.97$, the cross section $\sigma$ is
 constrained to be $
\lesssim 0.2$ pb at $95\%$ C.L. Interpreting $X^0$ as an
unparticle, bounds can be obtained for combinations of parameters
$\lambda_i$ and $\Lambda_\U$. The corresponding bounds on various
cross sections we are interested can then be obtained.
We list them in Table I for $d_\U = 1.5$.

No observation of $e^+ e^- \to \gamma + \U$ at LEP II may be due
to too low event number. With larger integrated luminosity at ILC,
unparticle effects may be observed. If the scale of unparticle
physics is close to the upper bound, with an integrated luminosity
of 100 $fb^{-1}$, the event numbers can reach more than $1.93
\times 10^{4}$ ($1.3\times 10^{3}$) and $1.07\times 10^4$
($6.19\times 10^2$) for $e^+ e^- \to \gamma (Z) + \U$ with
$\sqrt{s} = 200$ GeV and $\sqrt{s} = 500$ GeV, respectively. With
an integrated luminosity of 500 $fb^{-1}$, the event numbers would
be five times larger. The ILC would be able to study the detailed
distributions discussed earlier, and provide crucial information
on the properties of the unparticles.
\begin{table}
\begin{center}
\begin{tabular}{|c|c|c|c|c|}
\hline
& \multicolumn{2}{c|}{$\sqrt{s}=200$ GeV} & \multicolumn{2}{c|}{$\sqrt{s}=500$ GeV}\\
\cline{2-5}
  $d_\U=1.5$ & $\sigma(e^+e^-\to\gamma \U)$
  & $\sigma(e^+e^-\to Z \U)$ & $\sigma(e^+e^-\to\gamma \U)$ & $\sigma(e^+e^-\to Z \U)$  \\
\hline\hline
  $\lambda_{LL}$-term  & $0.193$ pb & $0.0410$ pb  & $0.483$   pb  & $0.482$ pb      \\
  $\lambda_{EE}$-term  & $0.193$ pb & $0.0266$ pb  & $0.483$   pb  & $0.313$ pb      \\
  $\lambda'_{LL}$-term & $0.204$ pb & $0.0133$ pb  & $0.107$   pb  & $0.00954$ pb      \\
  $\lambda'_{EE}$-term & $0.204$ pb & $0.00863$ pb & $0.107$   pb  & $0.00619$ pb      \\
  $\lambda_{ww} $-term      &  $0.198$ pb &  $0.369$   pb & $0.365$ pb &   $0.809$ pb    \\
  $\tilde\lambda_{ww}$-term  &  $0.198$ pb &  $ 0.0969$ pb  & $0.365$ pb& $0.616 $ pb   \\
  $\lambda_{bb} $-term       &  $0.195$ pb &  $0.162$   pb & $0.434$ pb &   $0.355$ pb    \\
  $\tilde\lambda_{bb}$-term  &  $0.195$ pb &  $ 0.0426$ pb  & $0.434$ pb& $0.270 $ pb   \\
  \hline
\end{tabular}
\caption{Bounds on the cross sections using LEP II constraints
with the cuts $|\cos\theta| < 0.97$ and $E_\gamma > 5$ GeV. Event
numbers can be obtained by multiplying a given integrated
luminosity.}
\end{center}
\end{table}

There are also several non-collider laboratory constraints
directly related to the operators have been studied including g-2
of electron\cite{Cheung:2007ue,Luo:2007bq,Liao:2007bx} and
invisible positronium decays\cite{Liao:2007bx}. We find that
positronium decays into an $\U$ may provide a stronger bound than
that at LEP II.

The operators with couplings $\lambda_{LL, EE}$ and $\lambda_{LL,
EE}^{\prime}$ contribute to para-positronium to unparticle
(p-Ps$\to \U$)) and ortho-Positronium to unparticle (o-Ps$\to \U$)
decays directly at the tree level, respectively. The 90\% C.L.
experimental bounds on these decay branching
ratios\cite{Ps-bound}, $4.3\times 10^{-7}$ and $4.2\times
10^{-7}$, then lead to \cite{footnote}
\begin{eqnarray}
&&A_{d_\U}|\lambda_{LL}-\lambda_{EE}|^2 (2m_e/\Lambda_\U)^{2d_\U}
< 9.2\times 10^{-9}\;,
\nonumber\\
&&A_{d_\U}|\lambda^{\prime}_{LL}+\lambda^{\prime}_{EE}|^2
(2m_e/\Lambda_\U)^{2(d_\U-1)} < 2.0\times 10^{-12}\;.
\end{eqnarray}
In the above, our bound on scalar unparticle scale is different
than that obtained in Ref.~\cite{Liao:2007bx} is due to the fact
that we have required the unparticles to couple to SM invariant
operators such that the couplings scale as $\Lambda_\U^{-2d_\U}$
while in Ref.~\cite{Liao:2007bx} they scale as
$\Lambda_\U^{-2(d_\U -1)}$.

The constraints on $\lambda_{LL,EE}$ are much weaker than that
from LEP II. The constraints on $\lambda^{\prime}_{LL,EE}$ are,
however, much stronger. Assuming no cancellation between
$\lambda^{\prime}_{LL}$ and $\lambda^{\prime}_{EE}$, the bounds
would imply that the cross sections for $e^+ e^-\to \gamma (Z) +
\U$ to be less than $5.5\times 10^{-5} (3.4\times 10^{-6})$ pb at
$\sqrt{s}=200$ GeV and $2.9\times 10^{-5} (2.3\times 10^{-6})$ pb
at $\sqrt{s}=500$ GeV for $d_\U = 1.5$. If true, the unparticle
effects due these operators will not be able to be studied at ILC.
However, with larger $d_\U$, it is still possible. For example,
for $d_\U =1.88$, one can get more than $4.6 \times 10^4
(1.2\times 10^3)$ and $5.1 \times 10^4 (2.6\times 10^3)$
$e^+e^-\to \gamma (Z) + \U$ events at $\sqrt{s}=200$ GeV and $500$
GeV, respectively with the integrated luminosity of $100 fb^{-1}$.
We should also note that since the constraint is proportional to
$\lambda_{LL}^{\prime} + \lambda^{\prime}_{EE}$, if there is a
cancellation such that this quantity is small, but individual
$\lambda^{\prime}_{LL,EE}$ is not small, the cross sections for
$e^+e^-\to \gamma (Z) + \U$ can still be large and unparticle
physics effects can still be studied at ILC.

No constraints on $\lambda_{ww, bb}(\tilde \lambda_{ww, bb}) $ can
be obtained from a positronium decays into an unparticle. However,
one can obtain constraints from o-Ps$\to \gamma + \U$ decay
through the diagrams shown in Fig. 1. We have carried out such
study using formulae in Ref.~\cite{Chen:2007zy} for $\Upsilon \to
\gamma + \U$ with appropriate replacement for parameters. We find
that the constraints are much weaker than that from LEP II data.

In summary, unparticle physics due to scale invariance leads to
very rich collider and flavor phenomenology. Under the scenario
that the unparticle stuff transforms as a singlet under the SM
gauge group, we listed possible operators involving interactions
of scalar $\U$, vector $\U^\mu$ and spinor $\U^s$ unparticles with
the SM fields and derivatives up to dimension four and discussed
some phenomenology related to these operators. We find that the
interactions of unparticle with Higgs sector and lepton sector are
quite interesting. We also find that $e^+e^-$ collider can provide
useful information for scalar and vector unparticles.

\vskip 1.0cm
\noindent {\bf Acknowledgments}$\,$ The work of authors was
supported in part by the NSC and NCTS.

\end{document}